\documentclass[floatfix,aps,prd,showpacs,amsmath,nofootinbib,preprintnumbers]
{revtex4}

\usepackage{graphicx}

\begin{document}

\preprint{FERMILAB-PUB-04-292-A}

\title{Curvature Perturbations from Broken Symmetries}

\author{Edward W. Kolb}\email{rocky@fnal.gov}
\affiliation{Particle Astrophysics Center, Fermi
        National Accelerator Laboratory, Batavia, Illinois \ 60510-0500, USA \\
        and Department of Astronomy and Astrophysics, Enrico Fermi Institute,
        University of Chicago, Chicago, Illinois \ 60637-1433, USA}

\author{Antonio Riotto}\email{antonio.riotto@pd.infn.it}
\affiliation{INFN, Sezione di Padova, via Marzolo 8, I-35131, Italy}

\author{Alberto Vallinotto}\email{avallino@uchicago.edu}
\affiliation{Physics Department, The University of Chicago,
    Chicago, Illinois 60637-1433, USA \\
    and Particle Astrophysics Center, Fermi
        National Accelerator Laboratory, Batavia, Illinois \ 60510-0500, USA}

\date{\today}

\begin{abstract}
We present a new general mechanism to generate curvature
perturbations after the end of the slow-roll phase of
inflation. Our model is based on the simple assumption
that the potential driving inflation is characterized by an
underlying global symmetry which is slightly broken.
\end{abstract}


\maketitle

\section{Introduction \label{sec:INTRO}}

One of the most successful predictions of the inflationary theory, the
current paradigm for understanding the evolution of the early
universe \cite{guth81}, is the redshifting of quantum fluctuations
of the field driving inflation -- the \textit{inflaton} -- beyond
the Hubble radius, leading to an imprint on the background
scalar (density) and tensor (gravitational waves) metric perturbations
\cite{lrreview,muk81,hawking82,starobinsky82,guth82,bardeen83,
Hu:2004xd,Mukhanov:1990me,Durrer:2004fx,Langlois:2004de,Brandenberger:2003vk}
that subsequently seeds structure formation.

For simplicity, most inflation models assume that there is only one scalar
field involved in the dynamics of inflation. This is also the case
when the mechanism of converting the energy driving inflation into
radiation is considered. In this work we point out a qualitatively
new effect that might arise if one relaxes the assumption of a
single dynamical field. In a multi-field scenario in which the
inflationary potential is characterized by a broken symmetry, the
quantum fluctuations generated during the inflationary stage
represent fluctuations in the initial conditions for the dynamics of
the inflaton in the subsequent stage, thus implying that the
background dynamics after the slow-roll phase has ended will differ
in different regions of the universe. Since the background fields
are coupled to the other fields into which they decay, the
fluctuations generated during the slow-roll phase will affect the
subsequent decay process.

The present work, assuming that the inflaton decay into other fields
through the non-perturbative process of \textit{preheating}
\cite{Kofman:1997yn,Felder:1998vq}, is then aimed to understand
whether isocurvature inflaton fluctuations, generated during the slow-roll
stage, can lead to perturbations of the background metric through variations of
the preheating efficiency. While the generation of curvature
perturbations during the stages following the slow-roll phase has
already been considered in some works
\cite{Dvali:2003em,Dvali:2003ar,Matarrese:2003tk,Lyth:2001nq,Liddle:1999hq,
Wands:2000dp,Malik:2002jb,Gupta:2003jc},
the present work is the first one to show that in a multi-field
scenario a global broken symmetry of the potential is sufficient to
yield curvature perturbations. Curvature perturbations produced
through this mechanism can even represent the main source of
perturbations to the background metric if the inflationary potential
is such that the mass required to produce quantum fluctuations along
the field trajectory is large, so that the latter result
exponentially suppressed.

The structure of the present work is the following. In Sec.\
\ref{sect:GI} we obtain a general formula for the curvature perturbations
generated from an inhomogeneous preheating efficiency related
to the quantum fluctuations produced during inflation.
Sec.\ \ref{sect:Application} presents an application of the general
result obtained in Sec.\ \ref{sect:GI} to the case of a broken
$U(1)$ symmetry. The conclusions are contained in Sec.\
\ref{sect:Discussion}.

\section{General Results \label{sect:GI}}

One of the main objectives in any particular preheating model is the
calculation of the comoving number density of particles produced
during the process, usually denoted by $n_{\chi}$. In general,
$n_{\chi}$ is a functional of the evolution and of the couplings of
the preheat field $\chi$ to the dynamically evolving field(s):
$n_{\chi}=F[\vec{\phi}(t)]$, where $\vec{\phi}= \phi_1,\ \phi_2,\
\cdots,\ \phi_n$ denotes the background inflaton fields that couple
to $\chi$. Choosing a specific preheating model is then equivalent
to specifying the functional $F$ that relates $\vec{\phi}(t)$ to
$n_{\chi}$.

In general, the dynamics of the background fields and of the scale
factor is given by the solution of the system of coupled
differential equations
\begin{subequations}
\begin{eqnarray}
    \ddot{\phi}_i+3H\dot{\phi}_i+\frac{\partial V}{\partial
    \phi_i}=0, \quad i=1,...,n,\label{GI:backdyn}\\
    H^2=\left(\frac{\dot{a}}{a}\right)^2=\frac{8\pi}{3 M_p^2}\left[
    \sum_i \frac{\dot{\phi}_i^2}{2}+V(\vec{\phi})\right],\label{GI:H2}
\end{eqnarray}
\end{subequations}
once the particular landscape of the potential $V(\vec{\phi})$ and
a set of initial conditions $\{ \vec{\phi}(t_0),
\dot{\vec{\phi}}(t_0) \}$ are specified.

The main subject of the present work is to analyze how quantum
fluctuations generated during the slow-roll stage of inflation may
affect the efficiency of the subsequent preheating process. As
inflation proceeds, quantum fluctuations will in fact cause the
value of the background inflaton field to fluctuate in space about a
mean value $\vec{\phi}_0(t)$:
\begin{equation}\label{GI:in_cond}
    \vec{\phi}(t,x)=\vec{\phi}_0(t)+\delta\vec{\phi}(t,x).
\end{equation}

Denoting by $t_0$ the epoch when inflation has ended but preheating
has not yet commenced, it is possible to note that Eq.
(\ref{GI:in_cond}) above shows that at each point in space the
initial conditions $\vec{\phi}(t_0,x)$ that will determine the
subsequent background field dynamics through Eqs.\
(\ref{GI:backdyn},\ref{GI:H2}) are affected by the quantum
fluctuations produced during inflation. Since the preheating
efficiency is related to the dynamics of the background, it is then
possible to conclude that quantum fluctuations produced during
inflation may lead to fluctuations in the preheating efficiency
through different background dynamics.

\begin{figure}
\includegraphics[width=0.65\textwidth]{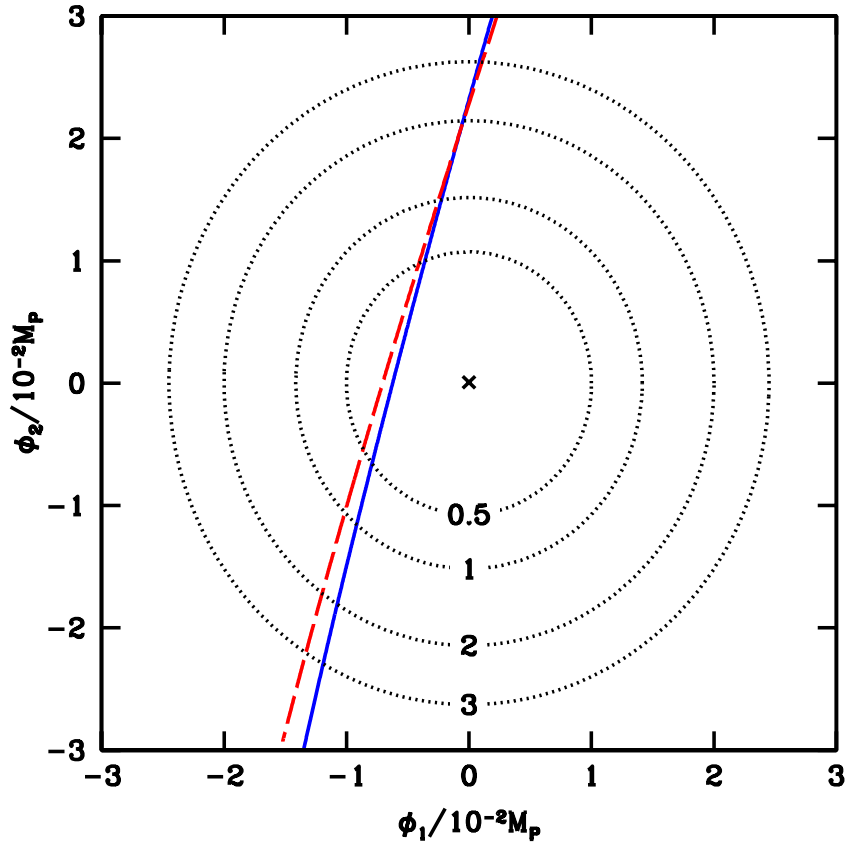}
\caption{\label{Fig:0}Two background trajectories arising from
fluctuations in the initial conditions in the case in which the
background field space is two dimensional. Also shown are the
equipotential contours (units of $10^{-16} M_p^4$). The potential
used in this figure is the one given in Eq.\ (\ref{App:V}) and
further analyzed in Sec.\ \ref{sect:Application}, which is
characterized by a global $U(1)$ broken symmetry. In this case, and
for illustration purposes only, the symmetry breaking parameter has
been arbitrarily set to $x=0.15$.}
\end{figure}

\textit{Broken Symmetry.} It is then necessary to point out that the
mere presence of quantum fluctuations in the initial conditions for
the dynamics of the background during the preheating stage are not
sufficient to yield different background evolutions leading to
fluctuations of the preheating efficiency. If in fact the background
potential $V(\vec{\phi})$ is perfectly symmetric -- that is
$V(\vec{\phi})=V(|\vec{\phi}|)$---then the fluctuations in the
initial conditions will only lead to background evolutions that are
time translations of each other: in this case a simple rotation of
the coordinate system in field space would again yield the well know
case of a single scalar field. If the inflationary potential is
characterized by a broken symmetry, on the other hand, then
fluctuations in the initial conditions will lead to background
trajectories that are not just time translations of one another. Two
such background trajectories are shown in Fig. \ref{Fig:0} for the
case of a two-dimensional field space.  Notice that the minimum distance
to the origin in the trajectories are different.  If the efficiency of
the preheating process depends on the minimum distance obtained in the
trajectory, then the preheating history will differ.

\textit{Initial Conditions.} It is important to distinguish here
between the ``initial conditions'' for the background dynamics
specified at the beginning of the slow-roll phase and those
specified at time $t_0$, that is once the slow-roll phase has
terminated and the preheating phase has not yet commenced. Since the
present work deals with the particle production during preheating,
the term ``initial conditions''---and their fluctuations---refers
here to the initial conditions for the preheating phase, that is to
$\vec{\phi}(t_0,x)$.

Considering the presence of the friction term $3H\dot{\phi}_i$ in
Eqs.\ (\ref{GI:backdyn}), it also seems reasonable to assume that
the background dynamics during the preheating stage is mostly
affected by the position that the background occupies in field space
at the beginning of such a phase. Since the comoving number density
of particles produced during preheating is a functional of the
background trajectory, it is then possible to conclude that
\begin{equation}\label{GI:nchioftheta}
    n_{\chi}=n_{\chi}[\vec{\phi}(t_0),...].
\end{equation}

Let's then turn to the preheating process and to the generation of
curvature perturbations. As a first approximation, let's assume that
preheating is complete and that the products of preheating are
therefore the only particles populating the universe when the
preheating stage has ended.\footnote{This assumption is quite
important: since the preheat field $\chi$ is the only component
present in the universe at the end of preheating, this automatically
ensures that only curvature perturbations are present at that point.
On the other hand, if the preheating process doesn't turn all the
energy initially stored in the $\phi$ field into the $\chi$ field,
isocurvature perturbations can also result.} Neglecting the possible
contributions stemming from non-adiabatic pressure perturbations
present during the preheating stage, an estimate for the curvature
perturbation $\zeta$ can be obtained considering the number density
perturbation,
\begin{equation}\label{GI:curvpert}
 \zeta\equiv\psi-H\frac{\delta\rho_{\chi}}{\dot{\rho}_{\chi}}
 \approx\alpha\frac{\delta n_{\chi}}{n_{\chi}},
\end{equation}
where the \textit{spatially flat gauge} has been assumed and the
proportionality constant $\alpha$ depends on the redshifting of the
particle produced. The above expression then allows to obtain an
estimate of the curvature perturbations produced during the
preheating stage induced by the fluctuations in the initial
conditions $\delta\vec{\phi}(t_0)$ present at the beginning of such
a stage because of the preceding inflationary stage. To proceed
further it is then necessary to note that the coordinate system
chosen to express the potential is not necessarily the one suited to
analyze the perturbations arising during inflation, since in this
coordinate system adiabatic and isocurvature perturbations are not
decoupled. Adiabatic perturbations are most commonly thought to be
the ones dominating the energy density during the inflationary
stage, while the isocurvature ones are usually considered not to
affect the energy density. Recalling the work of by Gordon
\textit{et al.} \cite{Gordon:2000hv}, it is also possible to note
that given the inflationary trajectory $\vec{\phi}(t)$, adiabatic
perturbations correspond to perturbations along the direction
tangent to the trajectory $\dot{\vec{\phi}}(t)$, while isocurvature
perturbations correspond to perturbations in the hyperplane
orthogonal to $\dot{\vec{\phi}}(t)$. This point can be intuitively
understood once Eq.\ (\ref{GI:backdyn}) is considered: since the
motion of the background is driven by the gradient of the potential,
it is reasonable to expect that the trajectory will be tangent to
the gradient. This in turn means that perturbations perpendicular to
the trajectory will necessarily be isocurvature perturbations since
they lie on equipotential hypersurfaces (thus perturbing the field
values but not the energy density). Perturbations along the
direction of the trajectory, on the other hand, being orthogonal to
equipotential hypersurfaces, will necessarily affect the value of
the energy density and therefore correspond to adiabatic
perturbations. To decouple adiabatic and isocurvature perturbation
it is then necessary to rotate the field space coordinate system so
that one of the unit vectors lies along the tangent to the
trajectory. Let's first of all define the unit vector
$\hat{u}_{\parallel}$ which points in the direction of field space
parallel to the (tangent to the) trajectory by
\begin{equation}
    \hat{u}_{\parallel}=\frac{\dot{\vec{\phi}}(t_0)}{|\dot{\vec{\phi}}(t_0)|},
\end{equation}
which allows to determine the component of the perturbation parallel
to the trajectory simply by
$\delta\phi_{\parallel}=\delta{\vec{\phi}}(t_0)\cdot\hat{u}_{\parallel}$.
Using the latter, it is then possible to define the unit vector
$\hat{u}_{\perp}$ which is orthogonal to the (tangent to the)
trajectory by
\begin{equation}
    \hat{u}_{\perp}=\frac{\delta\vec{\phi}(t_0)-
    \delta\phi_{\parallel}\hat{u}_{\parallel}}
    {|\delta\vec{\phi}(t_0)-\delta\phi_{\parallel}\hat{u}_{\parallel}|},
\end{equation}
which then leads to the component of the perturbation orthogonal to
the trajectory
$\delta\phi_{\perp}=\delta{\vec{\phi}}(t_0)\cdot\hat{u}_{\perp}$. A
new coordinate system in field space has then been defined, and the
perturbation $\delta\vec{\phi}(t_0,x)$ has been decomposed
accordingly:
\begin{equation}
    \delta\vec{\phi}(t_0,x)=\delta\phi_{\parallel}\hat{u}_{\parallel}
    +\delta\phi_{\perp}\hat{u}_{\perp}.
\end{equation}
In this new coordinate system $\delta\phi_{\parallel}$ represents an
adiabatic perturbation while $\delta\phi_{\perp}$ represents an
entropy perturbation. It is now quite intuitive to note that if the
potential is characterized by a broken symmetry, then while
adiabatic perturbations of the initial conditions
$\delta\phi_{\parallel}$ will lead to background trajectories that
are differing just by a time translation, entropy perturbations will
lead to trajectories that substantially differ from each other and
that will therefore produce variations of the preheating efficiency.

With this redefinition of the coordinate system of field space it is
then possible to estimate the variation of the comoving density of
particles produced during preheating due to fluctuations in the
initial conditions generated during inflation. Note in fact that
\begin{equation}\label{GI:deltanovern}
    \frac{\delta n_{\chi}}{n_{\chi}}=\frac{\partial \ln (n_{\chi})}{\partial
    \phi_{\parallel}}\delta\phi_{\parallel}+
    \frac{\partial \ln (n_{\chi})}{\partial
    \phi_{\perp}}\delta\phi_{\perp}=\nabla\ln(n_{\chi})
    \cdot\delta\vec{\phi}(t_0,x),
\end{equation}
but since perturbations of the initial conditions parallel to the
field velocity will simply lead to background evolutions that are
time translations of each other it is possible to conclude that
$\partial \ln (n_{\chi})/\partial \phi_{\parallel}=0$ and that the
mechanism under analysis is thus able to convert entropy
perturbations into adiabatic ones.\footnote{This fact is obviously
related to the assumption that the inflaton is supposed to
completely decay into -- and \textit{only} into -- the $\chi$
field.} Also, it is possible to envision models in which the
inflationary dynamics is such that the adiabatic perturbations
$\delta \phi_{\parallel}$ are exponentially suppressed. Combining
Eqs.\ (\ref{GI:curvpert}) and (\ref{GI:deltanovern}) it is therefore
possible to conclude that an estimate of the curvature perturbations
produced by the inhomogeneous preheating efficiency connected to
fluctuations in the background field dynamics originated during the
inflationary phase is given by
\begin{equation}\label{GI:zeta}
    \zeta\approx\alpha\frac{\partial\ln(n_{\chi})}{\partial\phi_{\perp}}
    \delta\phi_{\perp},
\end{equation}
where it is interesting to note that while the $\delta\phi_{\perp}$
factor is determined during the inflationary stage, the
$\alpha\left[{\partial\ln(n_{\chi})}/{\partial\phi_{\perp}}\right]$
factor is determined by the details assumed for the preheating
process (and the associated particle theory). The \textit{general}
conclusion that really seems worth stressing though is that on
rather general grounds this model allows the conversion of entropy
perturbations into curvature perturbations.

It seems important to stress once more that the ``initial
conditions'' that are considered in this work for the background
field dynamics are the initial condition that result once inflation
has ended (that is when $\ddot{a}$ becomes negative). This is
because what does affect the preheating efficiency is the evolution
history of the background $\phi(t)$ when it oscillates about the
minimum of its potential. The value of $\delta\phi_{\perp}$
appearing in Eq.\ (\ref{GI:zeta}) should then be evaluated at the
beginning of the oscillatory phase of the background, after the slow
roll phase has ended. Considering a specific $k-$mode, it is then
possible to note that the value of the amplitude of the quantum
fluctuations $\delta\phi_{\perp}(k)$ is frozen after the
corresponding wavelength has exited the horizon and that therefore
it can safely be evaluated when the wavelength $\lambda=2\pi/k$
crosses the horizon.\footnote{The calculation of the amplitude of
such quantum fluctuation, along with its power spectrum and the
resulting power spectrum and spectral index for the curvature
perturbation, is presented in the next section for the specific case
of a parabolic potential with a broken cylindrical symmetry.}
Recalling the general definition for the power spectrum of a generic
quantity $\delta\sigma$
\begin{equation}
    \mathcal{P}_{\delta\sigma}(k)\equiv\frac{k^3}{2\pi^2}|\delta\sigma_k|^2,
\end{equation}
it is  possible to see that the power spectrum and the spectral
index of the curvature perturbations obtained through this
mechanism are given by
\begin{subequations}
\begin{eqnarray}
    \mathcal{P}_{\zeta}(k)&=&\left[\alpha\frac{\partial\ln(n_{\chi})}
    {\partial\phi_{\perp}}
    \right]^2 \mathcal{P}_{\delta\phi_{\perp}}(k),\label{GI:powerspectrum}\\
    n-1&=&\frac{d\ln \mathcal{P}_{\zeta}}{d \ln k}=
    \frac{d\ln \mathcal{P}_{\delta\phi_{\perp}}}{d \ln k}.
    \label{GI:spectralindex}
\end{eqnarray}
\end{subequations}
From these expressions it is also important to point out that
while the power spectrum is affected by the specific nature of the
preheating process, the spectral index is affected only by the
characteristics of the potential in the region where quantum
fluctuations are stretched to superhorizon scales (which are
reflected in the power spectrum of $\delta\phi_{\perp}$).

\section{Application to the Broken $U(1)$ Case
\label{sect:Application}}

Let's apply the previous general results to the case in which the
scalar field landscape is described by two degrees of freedom,
$\phi_1$ and $\phi_2$. In this case, it is useful to express the
potential in terms of a complex field $\phi$,
\begin{equation}\label{GI:phicomplex}
    \phi=\phi_1+i\phi_2=|\phi|e^{i\theta}.
\end{equation}

If the potential $V(\phi_1,\phi_2)$ is characterized by an exact
$U(1)$ global symmetry, then at the end of inflation the
trajectory in field space will be in the radial
direction.\footnote{This is due to the fact that while the radial
acceleration has a source term from the potential, if there is a
$U(1)$ symmetry then the angular component has only the damping
term $3H\dot{\theta}$ arising from the expansion of the universe.
The presence of the damping term then causes any initial angular
velocity $\dot{\theta}_0$ to decay away.} In this case a simple
rotation of the coordinate system in field space would yield again
the well known case of a single scalar field, which then implies
that fluctuations in the angular component -- which in this case
corresponds to the previous $\hat{u}_{\perp}$ direction -- of the
initial conditions would not affect the background dynamics. Let's
then investigate what are the consequences on the preheating
process of an inflationary potential characterized by a slightly
broken $U(1)$ symmetry.

\subsection{Assumptions and Basic Results}

Following the notation of the previous section, the initial
conditions for the background field trajectory can be specified by
$[\phi_1(t_0), \phi_2(t_0)]$ or by $[|\phi_0|,\theta_0]$ (where,
since there is no possibility of confusion, the subscript $0$ here
refers to the initial conditions) and their corresponding time
derivatives. As was argued in the general case, the fluctuations in
the initial field velocities can be neglected. Furthermore,
recalling the presence of the $3H\dot{\phi}_i$ damping term in the
background equations of motion it is possible to argue that after a
first transient the trajectory in field space will be mostly along
the radial direction.\footnote{``Mostly'' because the symmetry
breaking term contribute a small source term to the angular
velocity.} It is therefore immediate to identify the new coordinate
system for the field space as
\begin{subequations}
\begin{eqnarray}
    \hat{u}_{\parallel}&\simeq&\widehat{|\phi_0|},\\
    \hat{u}_{\perp}&\simeq&\widehat{\theta_0}.
\end{eqnarray}
\end{subequations}

In the present case the comoving number density of particles
produced during preheating $n_{\chi}$ will therefore be a function
of the initial conditions:
$n_{\chi}=n_{\chi}[|\phi_0|,\theta_0,...]$. It is then possible to
apply Eq.\ (\ref{GI:zeta}) above to produce an estimate of the
curvature perturbations produced during the preheating stage caused
by the fluctuations in the angular direction $\delta\theta_0$
present at the beginning of such a stage,
\begin{equation}\label{App:deltanovern}
 \zeta\approx\alpha\frac{d\ln(n_{\chi})}{d\theta_0}|\delta\theta_0|.
\end{equation}

The power spectrum and the spectral index of the curvature
perturbation thus obtained can also be estimated applying Eqs.\
(\ref{GI:powerspectrum},\ref{GI:spectralindex}):
\begin{subequations}
\begin{eqnarray}
    \mathcal{P}_{\zeta}(k)&=&\left[\alpha\frac{d\ln(n_{\chi})}{d\theta_0}
    \right]^2 \mathcal{P}_{\delta\theta_0}(k),\label{Applic1:powerspectrum}\\
    n-1&=&\frac{d\ln \mathcal{P}_{\zeta}}{d \ln k}=
    \frac{d\ln \mathcal{P}_{\delta\theta_0}}{d \ln k}.
    \label{Applic1:spectralindex}
\end{eqnarray}
\end{subequations}

To proceed any further in the calculation it is necessary to
specify the two details that so far have been left completely
general. The first detail pertains the actual form of the
inflatonary potential. Since the $U(1)$ symmetry is assumed to be
slightly broken, we assume that it takes the simple form
\begin{equation}\label{App:V}
V(\phi_1,\phi_2)=\frac{m^2}{2}\left[\phi_1^2+\frac{\phi_2^2}{(1+x)}\right],
\end{equation}
where $x$ represents a measure of the symmetry
breaking.\footnote{Recalling the fact that during preheating the
value of $n_{\chi}$ is an adiabatic invariant and that it changes
only when the background field is located in a small region of field
space surrounding the the minimum of its potential, it is possible
to note that Eq.\ (\ref{App:V}) represents quite a general choice
since in such a region any potential can be well approximated in
this form. On the other hand the spectrum of the initial condition
perturbation $\delta\theta_0$ will depend on the form taken by the
potential in the region where it drives inflation. The specification
of Eq.\ (\ref{App:V}) above then can be considered general as far as
the estimation of the ${d\ln (n_{\chi})}/{d\theta_0}$ factor is
concerned, but it is not general at all once the estimate of
$\delta\theta_0(k)$ is considered.} The origin of a nonvanishing
value of $x$ may be gravitational effects which can strongly violate
global symmetries \cite{gravity}. In such a case, $x$ is likely to
be given by (some power of) the ratio of the fundamental energy
scale in the problem to the Planckian scale.

The second detail that needs to
be specified is the actual preheating model, thus nailing the
specific functional that connects the background dynamics to the
comoving number density of particles. In the present work the
preheating model assumed is the \textit{instant preheating} model of
Felder {\it et al.} \cite{Felder:1998vq}. It is in fact not so
unreasonable to suppose that the preheat field $\chi$ is coupled to
some other fields into which it can decay. Furthermore, this choice
for the preheating model is also characterized by some computational
simplicity since in this case it is possible to express $n_{\chi}$
as a function of the initial conditions imposed on the background
dynamics without having to resort to heavy numerical simulations.

The instant preheating model \cite{Felder:1998vq} assumes that the
inflaton field $\phi$ is coupled to the preheat field $\chi$
through the standard (and simplest) preheating interaction,
$\mathcal{L}_{\phi\chi}=-\frac{1}{2} g^2 |\phi| ^2 \chi ^2$, and
that the field $\chi$ is \textit{also} coupled to a fermion field
$\psi$ by the interaction
$\mathcal{L}_{\chi\psi}=h\bar{\psi}\psi\chi$.\footnote{We suppose
the fermion field $\psi$ to be massless, but this assumption is
not crucial.} Depending on the value of the coupling constants,
the process $\phi\rightarrow\chi\rightarrow\psi$ can be very
efficient and turn the energy density initially stored in the
background field into fermions in a single half oscillation of the
inflaton about the minimum of its potential.

Applying the results first obtained by Kofman {\it et al.}\
\cite{Kofman:1997yn}, it is possible to compute the comoving
number density of particles produced during the first pass of the
background inflaton about the minimum of the potential. Given the
interaction Lagrangian, it is important to note that if the
inflaton trajectory doesn't exactly pass through the minimum of
the potential (located at the origin of
the coordinates in field space) but at a minimum distance
$|\phi_*|$, then the preheat particles generated will be
characterized by an effective mass $m_{\chi}=g|\phi_*|$. The
comoving number density of $\chi$ particles produced in this case
is then given by \cite{Felder:1998vq,Kofman:1997yn}
\begin{equation}\label{App:n_chi}
    n_{\chi}=\frac{\left(g
    |\dot{\phi}_*|\right)^{3/2}}{8\pi^3}\exp\left[ -\frac{\pi
    g|\phi_*|^2}{|\dot{\phi}_*|}\right],
\end{equation}
where $t_*$ denotes the instant in which the inflaton reaches the
minimum of the potential $V$ \textit{along its trajectory} and
$|\dot{\phi}_*|$ and $|\phi_*|$ respectively denote the field
velocity and distance from the origin at such an instant.

\subsection{Estimate of the Curvature Perturbations}

Estimation of the curvature perturbations are not so complicated.
Given the interaction Lagrangian connecting the preheat field
$\chi$ with the fermion field $\psi$, the decay rate for the
perturbative process $\chi\rightarrow\psi\bar{\psi}$ is given by
$\Gamma_{\chi\rightarrow\psi\bar{\psi}}=h^2m_{\chi}/8\pi=h^2g|\phi|/8\pi$.
Since the decay rate for this process increases as the inflaton
moves away from the minimum, it is not so unreasonable to assume
that the whole process $\phi\rightarrow\chi\rightarrow\psi$ may be
completed in a single half oscillation and that at the end only
the fermions $\psi$ are going to be present. In the
\textit{spatially flat} gauge, the curvature perturbations will
then be given by
\begin{equation}\label{App:zeta2}
    \zeta\approx-H\frac{\delta\rho_{\psi}}{\dot{\rho_{\psi}}},
\end{equation}
where $\rho_{\psi}$ is the energy density of the fermion field.

Recalling the form of the interaction Lagrangian that connects the
inflaton and the preheat fields, it is also possible to note that
when the inflaton moves away from the minimum of its potential it
will endow the preheat field with an effective mass
$m_{\chi}=g|\phi|$ which will quickly grow, rendering the preheat
field nonrelativistic. Its energy density will therefore be given
by $\rho_{\chi}\approx n_{\chi}m_{\chi}=n_{\chi}g|\phi|$. As a
first approximation, let's suppose that all the $\chi$ particles
decay in one single instant $t_1$ when the background inflaton
field value is $|\phi(t_1)|$. Then
\begin{equation}\label{App:rhos}
    \rho_{\psi}(t_1)=\rho_{\chi}(t_1)=n_{\chi}g|\phi(t_1)|,
\end{equation}
which in turn implies
\begin{equation}\label{App:zeta3}
    \zeta\approx-H\frac{\delta\rho_{\psi}}{\dot{\rho}_{\psi}}\approx
    \alpha\frac{\delta n_{\chi}}{n_{\chi}},
\end{equation}
where the constant $\alpha$ depends on the redshift of the $\psi$
particles. In the present case, the $\psi$ particles are assumed to
be massless so $\alpha=1/4$. From Eq.\ (\ref{App:n_chi}), it is then
straightforward to compute
\begin{equation}\label{App:deltanchinchi}
    \frac{\delta n_{\chi}}{n_{\chi}}=\left(\frac{3}{2}+
    \frac{\pi g |\phi_*|^2}{|\dot{\phi}_*|}
    \right)\frac{\delta |\dot{\phi}_*|}{|\dot{\phi}_*|}-
    \frac{2\pi g |\phi_*|^2}{|\dot{\phi}_*|}\frac{\delta |\phi_*|}{|\phi_*|} .
\end{equation}

To connect this expression with fluctuations of the initial
conditions in the angular direction it is then necessary to
express $|\phi_*|$ and $|\dot{\phi}_*|$ as functions of the
initial conditions. This is not such a complicated task given the
fact that if we may neglect the expansion of the universe on the
time scale of the single oscillation during which preheating
occurs, we may exactly solve for the background dynamics (for more
details, see Appendix \ref{sect:Appendix}). Expressing the initial
conditions in polar coordinates as $[|\phi_0|,\theta_0]$, the two
parameters are approximately given by
\begin{subequations}
\begin{eqnarray}
    |\phi_*(|\phi_0|,\theta_0;x)|&\approx& \frac{|\phi_0| \pi x}{2\sqrt{2}}
    |\sin(2\theta_0)| ,
    \label{App:lapprox}\\
    |\dot{\phi}_*(|\phi_0|,\theta_0;x)|&\approx&
    m|\phi_0|\sqrt{1-x\sin^2(\theta_0)
    \label{App:vapprox}}.
\end{eqnarray}
\end{subequations}
As an intuitive check, it is possible to note that as $x\rightarrow
0$ and the symmetry is unbroken, Eqs.\
(\ref{App:lapprox},\ref{App:vapprox}) yield the correct asymptotic
behavior: $|\phi_*|$ vanishes because all background trajectories
pass through the origin, and the $\theta_0$ dependence of
$|\dot{\phi}_*|$ also disappears.

Given Eqs.\ (\ref{App:lapprox},\ref{App:vapprox}), it is
straightforward to compute their relative variations as a function
of the variation in the initial condition of the angle
$\delta\theta_0$:
\begin{subequations}
\begin{eqnarray}
    \frac{\delta |\phi_*|}{|\phi_*|}&=& 2\frac{\cos(2\theta_0)}{\sin(2\theta_0)}
    \delta\theta_0=2\cot(2\theta_0)\delta\theta_0 ,\\
    \frac{\delta
    |\dot{\phi}_*|}{|\dot{\phi}_*|}&=&-\frac{x}{2}\frac{\sin(2\theta_0)}
    {[1-x\sin^2(\theta_0)]}\delta\theta_0.
\end{eqnarray}
\end{subequations}
Use of the above expressions in Eq.\ (\ref{App:deltanchinchi}) then yields
\begin{eqnarray}
\frac{\delta n_{\chi}}{n_{\chi}} & = &f(\theta_0)\delta\theta_0
    =-\Bigg[\left(\frac{3}{2}+\frac{\pi g |\phi_*|^2}{|\dot{\phi}_*|}
    \right)\frac{x\sin(2\theta_0)}{2[1-x\sin^2(\theta_0)]}
    +\frac{4\pi g
    |\phi_*|^2}{|\dot{\phi}_*|}\cot(2\theta_0)\Bigg]\delta\theta_0 \nonumber\\
    & = & -x\frac{\sin(2\theta_0)}{\sqrt{1-x\sin^2(\theta_0)}}\Bigg\{
    \frac{3}{4\sqrt{1-x\sin^2(\theta_0)}}
    +\frac{\pi^3 |\phi_0| g x }{8m}\left[
    \frac{x\sin^2(2\theta_0)}{2[1-x\sin^2(\theta_0)]}
    +4\cos(2\theta_0)\right]\Bigg\}\delta\theta_0\label{App:deltanchinchi2} .
\end{eqnarray}

\begin{figure}
\includegraphics[width=0.65\textwidth]{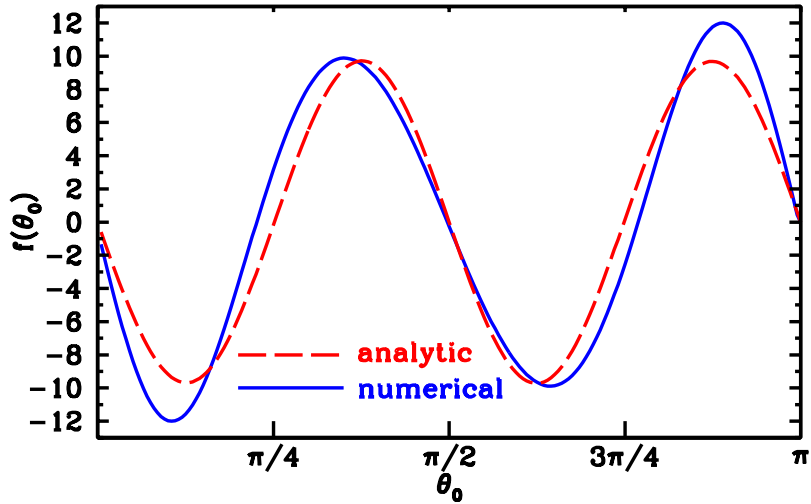}
\caption{\label{Fig:1}Values of $f(\theta_0)$ obtained through
numerical simulation and using the approximate expression
given in Eq.\ (\ref{App:deltanchinchi2}) for a value of the
symmetry breaking parameter of $x=0.05$ and for a coupling
constant $g=0.01$.}
\end{figure}

It is interesting to note that while the second and third term in
the curly brackets are coming from the exponential present in Eq.\
(\ref{App:n_chi}), the first one comes from the multiplicative term.
This means that for very small values of the symmetry-breaking
parameter $x$ the exponential suppression appearing in Eq.\
(\ref{App:n_chi}) does not apply since the trajectories of the
background field all pass extremely close to the origin of the
coordinate system and the $\chi$ particles generated are almost
massless.  For larger values of the symmetry-breaking parameter the
exponential suppression sets in and therefore the last two terms
become crucial in determining $\delta n_{\chi}/n_{\chi}$ because
small variation in the angle can lead to significant variations in
the suppressing exponential. A plot comparing the expression
obtained for $f(\theta_0)$ with the results of a numerical
simulation is given in Fig.\ \ref{Fig:1}. The final expression for
the curvature perturbations $\zeta$ produced through this mechanism
is thus given by
\begin{equation}\label{App:zetafin}
    \zeta=-\frac{x \sin(2\theta_0)}{4\sqrt{1-x\sin^2(\theta_0)}}\Bigg\{
    \frac{3}{4\sqrt{1-x\sin^2(\theta_0)}}
    +\frac{\pi^3 |\phi_0| g x }{8m}\left[
    \frac{x\sin^2(2\theta_0)}{2[1-x\sin^2(\theta_0)]}
    +4\cos(2\theta_0)\right]\Bigg\}\delta\theta_0,
\end{equation}
where it is furthermore possible to note that in the $x\rightarrow
0$ limit $\zeta$ vanishes. This fact is consistent with the point
raised above regarding the case of a perfect symmetry: if the
symmetry is unbroken then entropy perturbations will only lead to
background evolutions that are time translations of one another and
therefore the term ${d \ln (n_{\chi})}/{d \theta_0}$ vanishes
because $n_{\chi}$ doesn't depend on the angular initial condition.
Since $x$ plays a crucial role in
determining the overall scale of the density perturbations, it
would be interesting to investigate its value in realistic
models, {\it e.g.} in those cases  in which the breaking of the
global symmetry is due to gravitational effects.

Let's then proceed to compute the power spectrum and the spectral
index of the curvature perturbations. It has already been argued
that the amplitude $|\delta\theta_0(k)|$ can be evaluated at
horizon exit. Letting $H_k$ be the value of the Hubble parameter
when the wavelength $\lambda=2\pi/k$ crossed the Hubble radius during
inflation, it is
possible to show that for the potential of Eq.\ (\ref{App:V}) the
square of the amplitude of the quantum fluctuations is given by
\begin{equation}\label{App:deltatheta01}
    |\delta\theta_0(k)|^2=\frac{H_k^2}{2k^3|\phi|^2}
    \left(\frac{k}{aH}\right)^{2\eta_1}
    \left[1-2x\eta_1\cos^2(\theta_0)\ln\left(\frac{k}{aH}\right)\right],
\end{equation}
where $\eta_1=m^2/3H^2\ll 1$. Eq.\ (\ref{App:deltatheta01})
 shows that the symmetry breaking induces an very small
correction to the ordinary flat power spectrum
\begin{equation}
    \mathcal{P}_{\delta\theta_0}=\left(\frac{H_k}{2\pi|\phi|}\right)^2
    \left(\frac{k}{aH}\right)^{2\eta_1}\left[1-2\eta_1 x
    \cos^2(\theta_0)\ln\left(\frac{k}{aH}\right)\right].
\end{equation}
Now note that the factor $f(\theta_0)$ is uniquely determined by
the initial conditions on the angle and by the specification of
the potential. The power spectrum of the curvature perturbations
can then be obtained by
\begin{equation}
    \mathcal{P}_{\zeta}(k) = \frac{f^2(\theta_0)}{16}
    \mathcal{P}_{\delta\theta_0}(k)
    =\frac{f^2(\theta_0)}{16}\left(\frac{H_k}{2\pi|\phi|}\right)^2
    \left(\frac{k}{aH}\right)^{2\eta_1}\left[1-2\eta_1 x
    \cos^2(\theta_0)\ln\left(\frac{k}{aH}\right)\right]
    \approx \frac{f^2(\theta_0)}{16}\left(\frac{H_k}{2\pi|\phi|}\right)
    \label{App:Pzeta},
\end{equation}
where the explicit form obtained for $f(\theta_0)$ in this case
[Eq.\ (\ref{App:deltanchinchi2})] has not been entered for sake of
brevity. Eq.\ (\ref{App:Pzeta})  then shows that the power
spectrum of the curvature perturbations generated through this
process is flat to a very good degree. This last aspect can be
further stressed by the calculation of the spectral index, which
gives
\begin{equation}
    n-1 = \frac{d\ln\mathcal{P}_{\zeta}}{d\ln k}=2\eta_1
    -\frac{2\eta_1x\cos^2(\theta_0)}{1-2\eta_1 x \cos^2(\theta_0)
    \ln(k/aH)}
    \approx 2\left[1-x\cos^2(\theta_0)\right]\frac{m^2}{3H^2},
\end{equation}
thus showing that a very small tilt is induced by the symmetry
breaking of the potential and the angle of the background
trajectory.

\section{Discussion \label{sect:Discussion}}

The analysis of this paper suggests that the production of
curvature perturbations due to a broken symmetry of the
inflationary potential and the resulting inhomogeneous efficiency
of the preheating stage may be a rather common phenomenon. It is,
in fact, appropriate to stress that while the magnitude of the
curvature perturbations produced through this mechanism will
depend on the details chosen for the specific model (potentials,
coupling constants, interaction Lagrangians, preheating mechanism,
and so forth), the mere fact that the inflationary potential has a
minimum characterized by a broken symmetry is sufficient to
guarantee the generation of curvature perturbations during the
preheating phase. This is because perturbations in a direction
orthogonal to the field trajectory yield evolutions of the
background that are not just time translations of one other (as
would be the case with perturbations along the direction of the
trajectory) but that might differ substantially. Such different
background evolutions then necessarily lead to different
preheating efficiencies, thus resulting in perturbations in the
comoving particle number and energy densities.

As it has been shown in Sec.\ \ref{sect:Application}, choosing a
specific preheating model allows one to quantify the magnitude of
the curvature perturbations produced by this mechanism and to
assess whether these may or may not represent a dominant component
with respect to the adiabatic perturbation produced during the
slow-roll phase by fluctuations along the radial direction. In the
present context, the choice of the \textit{instant preheating}
model of Felder {\it et al.}\ \cite{Felder:1998vq} has been made because
it seems plausible that the preheat field $\chi$ may be
coupled to some other fields. Moreover, the nature of such process
allows one to obtain convenient analytic estimates of the comoving
number density of particles produced that are not affected by the
stochasticity, related to the build up of preheat particles,
usually present in the standard preheating models
\cite{Kofman:1997yn}. Nonetheless, it seems important to stress
that the main conclusions of this work do not depend on the choice
of the preheating process, but only on the fact that $n_{\chi}$
depends on the background history, which in turn depends on the
initial conditions.

Finally, it seems important to note that if the preheat field is
coupled to, and decays into, one single field, then the effect of a
broken symmetry of the inflationary potential is to convert
isocurvature perturbations into adiabatic perturbations
\cite{Gordon:2000hv}. In this sense, the above model resembles in
spirit the curvaton model of Lyth and Wands \cite{Lyth:2001nq}, but
does not require the assumption of an external field.

\acknowledgments{}
E.W.K.\ and A.V.\ were supported in part by NASA grant NAG5-10842
and by the Department of Energy.

\appendix

\section{Approximate values of
$|\phi_*(|\phi_0|,\theta_0;x)|$ and
$|\dot{\phi}_*(|\phi_0|,\theta_0;x)|$ \label{sect:Appendix}}

\begin{figure}
\includegraphics[width=0.65\textwidth]{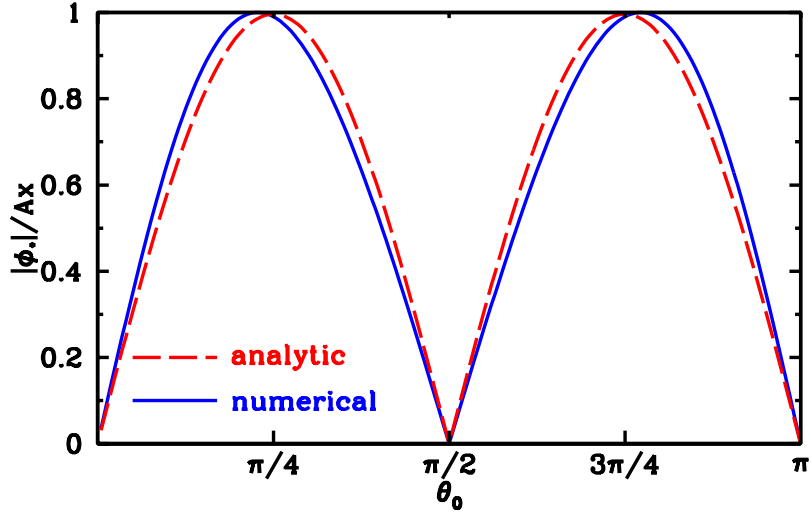}
\caption{\label{fig:Append:1}Values of $|\phi_*|/(|\phi_0|\pi x
/2\sqrt{2})$ obtained through numerical simulation and using the
above approximate expression Eq.\ (\ref{Append:lapprox}) for a value
of the symmetry breaking parameter of $x=0.05$.}
\end{figure}

First consider Eq.\ (\ref{App:lapprox}). It is straightforward to
note that given the inflationary potential, Eq.\ (\ref{App:V}),
the exact solution for the background dynamics can be computed
once the expansion of the Universe is neglected. The background
field dynamics is then given by
\begin{subequations}
\begin{eqnarray}
  \phi_1(t) &=& |\phi_0| \cos(\theta_0)\cos(m t), \\
  \phi_2(t) &=& |\phi_0| \sin(\theta_0)\cos\left(\frac{m t}{\sqrt{1+x}}\right) .
\end{eqnarray}
\end{subequations}
Letting $\Delta=\pi x/4$, it is possible to linearize the trajectory by
\begin{subequations}
\begin{eqnarray}
  \phi_1(t) &\approx& -|\phi_0| \cos(\theta_0) \sin(\Delta) \lambda \\
  \phi_2(t) &\approx& |\phi_0| \sin(\theta_0) \sin(\Delta) (1-\lambda) ,
\end{eqnarray}
\end{subequations}
with $\lambda\in[0,1]$. Then solving for the minimum distance from
the origin yields $\lambda=\sin^2(\theta_0)$, which
corresponds to a minimum distance $|\phi_*|$ of
\begin{equation}\label{Append:lapprox}
    |\phi_*|=\frac{|\phi_0| \pi x}{2\sqrt{2}}\sin(2\theta_0),
\end{equation}
where the approximation $\sin(\Delta)\approx\Delta$ has also been
used. A plot comparing the analytical approximation obtained in this
way with a numerical simulation is shown in Fig.\ \ref{fig:Append:1}.

Not let's then turn to Eq.\ (\ref{App:vapprox}). It is a basic fact of
a simple harmonic oscillator that $|\dot{\phi}_*|=m\Phi$ where
$\Phi$ is the amplitude of the oscillation and the mass $m$ is
given by the second derivative of the potential along the
trajectory
\begin{equation}\label{Append:m1}
    m^2=\frac{d^2V}{dl^2}.
\end{equation}
Here, $l$ parametrizes the trajectory. To obtain a good estimate for
$|\dot{\phi}_*|$ let's first assume that all trajectories  pass
through the origin. We define
\begin{subequations}
\begin{eqnarray}
  m_1^2 &=& \frac{\partial^2 V}{\partial \phi_1^2}=m^2, \\
  m_2^2 &=& \frac{\partial^2 V}{\partial \phi_2^2}=\frac{m^2}{1+x}.
\end{eqnarray}
\end{subequations}
If the trajectory makes an angle $\beta$ with the $\phi_1$ axis,
it is then straightforward to show that
\begin{equation}
    m_{\beta}^2=m_1^2 \cos^2(\beta)+m_2^2 \sin^2(\beta)
    =m^2\left[\cos^2(\beta)+\frac{\sin^2(\beta)}{1+x}\right] ,
\end{equation}
which for small value of $x$ reduces to
\begin{equation}\label{Append:m3}
    m_{\beta}\approx m \sqrt{1-x\sin^2{\beta}}.
\end{equation}

\begin{figure}
\includegraphics[width=0.65\textwidth]{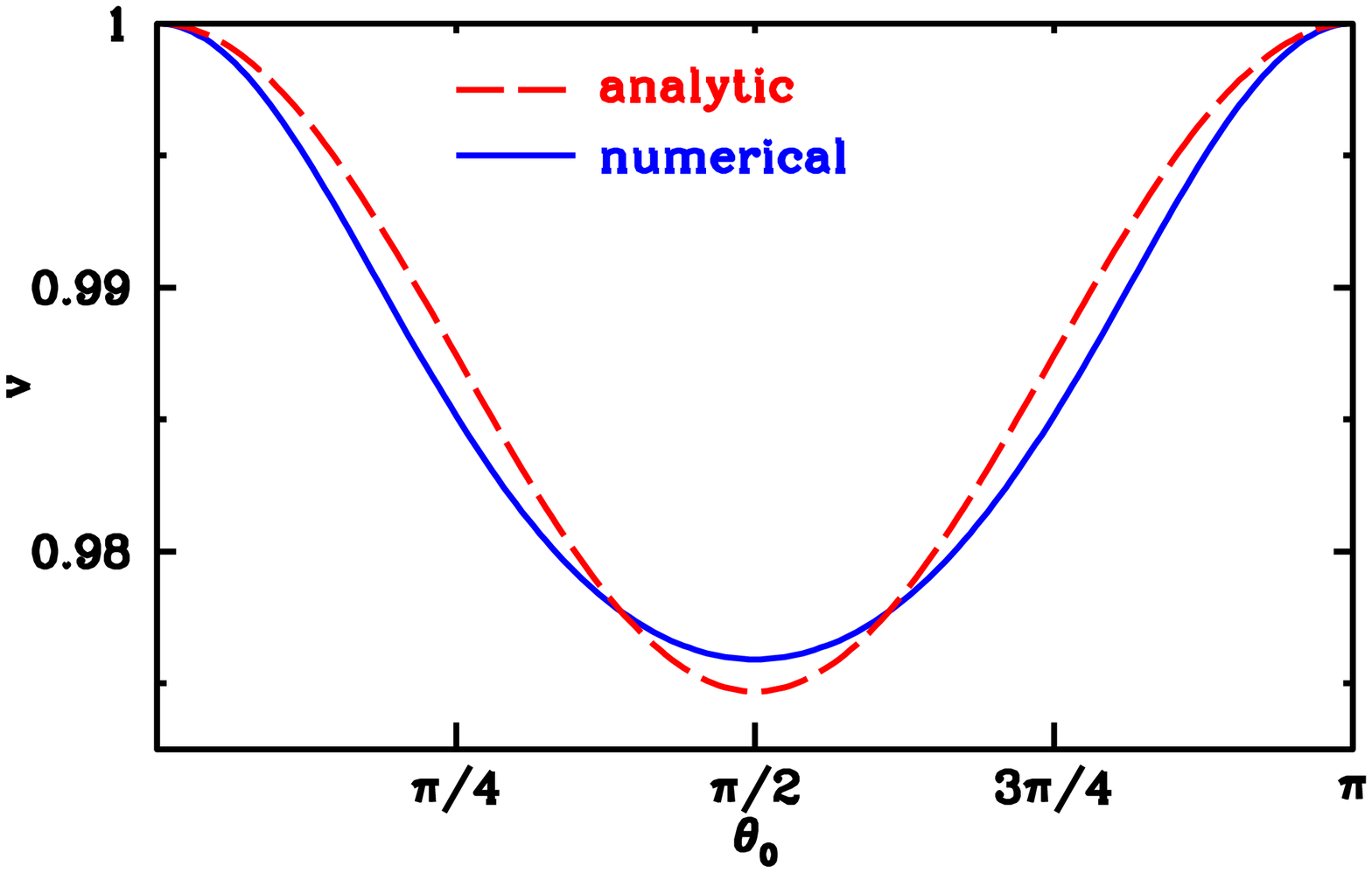}
\caption{\label{fig:Append:2}Values of
$|\dot{\phi}_*|/|\dot{\tilde{\phi}}_*|$ obtained through numerical
simulation and using the approximate expression Eq.\
(\ref{Append:m3}) for a value of the symmetry breaking parameter
of $x=0.05$.}
\end{figure}

Assuming that the trajectory is perfectly radial, which is
correct to a good approximation as long as the symmetry breaking
parameter $x$ is not too large, then $\beta=\theta_0$, and
therefore
\begin{equation}\label{Append:phidot}
    |\dot{\phi}_*|\approx |\dot{\tilde{\phi}}_*|\sqrt{1-x\sin^2{\theta_0}},
\end{equation}
where $|\dot{\tilde{\phi}}_*|=m|\phi_0|$ depends only on the initial
radial condition.\footnote{The fact that
$|\dot{\tilde{\phi}}_*|=m|\phi_0|$ simply follows from applying
conservation of energy to the background dynamics neglecting the
expansion of the Universe term.} A plot comparing the analytical
approximation given by Eq.\ (\ref{Append:m3}) with numerical
simulation is shown in Fig.\ \ref{fig:Append:2}.

\section{Calculation of the Power Spectrum and the Spectral Index}

Start from the fact that
\begin{eqnarray}
    \theta&=&\tan^{-1}\left(\frac{\phi_2}{\phi_1}\right) \label{App:theta}\\
    \Rightarrow
    \delta\theta_0&=&\sqrt{\left(\frac{\partial \theta}
    {\partial \phi_1}\right)^2\delta\phi_1^2
    +\left(\frac{\partial \theta}{\partial
    \phi_2}\right)^2\delta\phi_2^2},\label{App:deltatheta}
\end{eqnarray}
where the two scalar fields are supposed to be uncorrelated. Then
for a generic massive scalar field $\sigma_k$ we know that on
superhorizon scales the amplitude of the quantum fluctuations is
given by
\begin{equation}\label{App:deltachi}
    |\delta\sigma_k|\simeq\frac{H_k}{\sqrt{2k^3}}
    \left(\frac{k}{aH}\right)^{\frac{3}{2}-\nu_{\sigma}},
\end{equation}
where $\nu_{\sigma}^2=9/4-m_{\sigma}^2/H^2$ and it is possible to
define the parameter $\eta_{\sigma}=m_{\sigma}^2/3H^2 \simeq
3/2-\nu_{\sigma}$. In our case we have two fields, with masses
$m_{\phi_1}=m$ and $m_{\phi_2}=m/\sqrt{1+x}$. We can then use Eq.\
(\ref{App:deltachi}) to calculate $|\delta\theta_0(k)|$ using Eq.\
(\ref{App:deltatheta}). However since we're really interested in
computing the power spectrum, we can go directly to its formula,
which requires $|\delta\theta_0(k)|^2$. In general, the power
spectrum is defined by
\begin{equation}
    \mathcal{P}_{\delta\sigma}\equiv\frac{k^3}{2\pi^2}|\delta\sigma_{k}|^2,
\end{equation}
so that the first goal of our calculation becomes
\begin{equation}
    \mathcal{P}_{\delta\theta}\equiv\frac{k^3}{2\pi^2}|\delta\theta_0(k)|^2,
\end{equation}
but using Eqs.\ (\ref{App:theta}, \ref{App:deltatheta}) above we
then have
\begin{equation}
    |\delta\theta_0(k)|^2=\frac{\phi_2^2}{|\phi|^4}|\delta\phi_1|^2+
    \frac{\phi_1^2}{|\phi|^4}|\delta\phi_2|^2.
\end{equation}

Now the only problem is that since the two fields have slightly
different masses, we don't have $|\delta\phi_1|=|\delta\phi_2|$.
Instead
\begin{equation}
    |\delta\theta_0(k)|^2=\frac{H_k^2}{2k^3|\phi|^2}
    \left[\sin^2(\theta_0)\left(\frac{k}{aH}\right)^{2\eta_1}+
    \cos^2(\theta_0)\left(\frac{k}{aH}\right)^{2\eta_2}\right],
\end{equation}
but
\begin{equation}
    \eta_2=\frac{m^2}{3H^2(1+x)}\approx\frac{m^2}{3H^2}(1-x)=\eta_1(1-x),
\end{equation}
so that
\begin{equation}
    |\delta\theta_0(k)|^2=\frac{H_k^2}{2k^3|\phi|^2}
    \left(\frac{k}{aH}\right)^{2\eta_1}
    \left[\sin^2(\theta_0)+\cos^2(\theta_0)
    \left(\frac{k}{aH}\right)^{-2x\eta_1}\right].
\end{equation}
But recall that $x$ is very small, so that we can use the fact
that $a^x\approx 1+x\ln(a)$ which then yields
\begin{equation}
    |\delta\theta_0(k)|^2\approx\frac{H_k^2}{2k^3|\phi|^2}
    \left(\frac{k}{aH}\right)^{2\eta_1}
    \left[1-2x\eta_1\cos^2(\theta_0)\ln\left(\frac{k}{aH}\right)\right].
\end{equation}

We're now ready to insert this expression in the general
definition of the power spectrum and then we get:
\begin{eqnarray}
    \mathcal{P}_{\delta\theta}&=&\frac{k^3}{2\pi^2}\frac{H_k^2}{2k^3|\phi^2|}
    \left(\frac{k}{aH}\right)^{2\eta_1}\left[1-2\eta_1 x
    \beta^2\ln\left(\frac{k}{aH}\right)\right]\nonumber\\
    &=&\left(\frac{H_k}{2\pi|\phi|}\right)^2
    \left(\frac{k}{aH}\right)^{2\eta_1}\left[1-2\eta_1 x
    \beta^2\ln\left(\frac{k}{aH}\right)\right],
\end{eqnarray}

We can then note that since $\cos^2(\theta_0)\in[0,1]$ and $x\ll 1$
the spectrum we get is almost flat. Furthermore, the spectral index
is given by
\begin{eqnarray}
    n-1&=&\frac{d\ln\mathcal{P}_{\zeta}}{d\ln k}=2\eta_1-
    \frac{2\eta_1x\cos^2(\theta_0)}{1-2\eta_1 x
    \cos^2(\theta_0)
    \ln(k/aH)}\nonumber\\
    &\approx& 2\eta_1\left[1-x\cos^2(\theta_0)\right]=
    2\left[1-x\cos^2(\theta_0)\right]\frac{m^2}{3H^2},
\end{eqnarray}
where the denominator has been approximated to one since the
$\eta_1 x\cos^2(\theta_0)$ term should be very small.



\end{document}